\documentclass[aps,showpacs,amsmath,amssymb,pra,superscriptaddress,twocolumn,floatfix]{revtex4-1}

\usepackage{graphicx}
\usepackage{amsmath}
\usepackage{hyperref}
\usepackage{dcolumn}
\usepackage{bm}
\usepackage{color}

\usepackage[utf8]{inputenc}

\allowdisplaybreaks[1]

\begin{document}

\title{Local symmetries and perfect transmission in aperiodic photonic multilayers}

\author{P.A. Kalozoumis}
\affiliation{Department of Physics, University of Athens, GR-15771 Athens, Greece}

\author{C. Morfonios}
\affiliation{Zentrum f\"ur Optische Quantentechnologien, Universit\"{a}t Hamburg, Luruper Chaussee 149, 22761 Hamburg, Germany}

\author{N. Palaiodimopoulos}
\affiliation{Department of Physics, University of Athens, GR-15771 Athens, Greece}

\author{F.K. Diakonos}
\affiliation{Department of Physics, University of Athens, GR-15771 Athens, Greece}

\author{P. Schmelcher}
\affiliation{Zentrum f\"ur Optische Quantentechnologien, Universit\"{a}t Hamburg, Luruper Chaussee 149, 22761 Hamburg, Germany}
\affiliation{The Hamburg Centre for Ultrafast Imaging, Universit\"{a}t Hamburg, Luruper Chaussee 149, 22761 Hamburg, Germany}

\date{\today}

\begin{abstract}

We develop a classification of perfectly transmitting resonances occuring in effectively one-dimensional optical media which are decomposable into locally reflection symmetric parts.
The local symmetries of the medium are shown to yield piecewise translation-invariant quantities, which are used to distinguish resonances  with arbitrary field profile from resonances following the medium symmetries.
Focusing on light scattering in aperiodic multilayer structures, we demonstrate this classification for representative setups, providing insight into the origin of perfect transmission.
We further show how local symmetries can be utilized for the design of optical devices with perfect transmission at prescribed energies.
Providing a link between resonant scattering and local symmetries of the underlying medium, the proposed approach may contribute to the understanding of optical response in complex systems.

\end{abstract}

\pacs{42.25.Hz  % Interference, optical
      78.67.Pt, % Optical properties of multilayers / of photonic structures
      78.67.Bf, % Optical properties of nanoscale materials and structures / nanocrystals
      78.20.Ci, % Transmission coefficients, optical
      }

\maketitle

\section{Introduction \label{introduction}} 

Transmission properties and their control in inhomogeneous media with complex geometric structure has developed into a field of intense study with applications to electronic \cite{PhysRevLett.91.108101,Macia2006,PhysRevB.76.235113}, photonic \cite{Hsueh2011_josab-28-11-2584,Werchner2009,Macia2012}, acoustic \cite{Hladky2013} and magnonic \cite{Hsueh2013} systems. 
Aperiodic systems possess a central role both in understanding the fundamental concepts which govern the transitions from perfectly periodic order to randomness, and in the development and design of devices with controllable transport properties. Photonic multilayered devices constitute a wide class of such systems, offering the unique possibility to relate geometrical with optical properties in a direct and efficient manner. 
Typical examples are photonic multilayers possessing a quasiperiodic Fibonacci \cite{Kohmoto1987,Gellermann1994,Macia1998}, fractal Cantor \cite{Sun1991,Lavrinenko2002,Zhukovsky2004,Zhukovsky2005,Chuprikov2006} or even more general aperiodic geometry \cite{Zhukovsky2008,Nava2009,Zhukovsky2010}, leading to scaling and self-similarity of the corresponding optical transmission spectra.
Of particular interest is the case of perfect (that is, reflectionless) light transmission through an aperiodic multilayer.
There are many cases supporting that the presence of mirror symmetry in a multilayer device leads to perfectly transmitting resonances (PTRs), while the lack of such symmetry is usually accompanied by non-vanishing reflection. 
An instructive example of this scenario is the occurrence of PTRs in multilayers with Fibonacci order after appropriate symmetrization of the device \cite{Huang1999,Huang2001,Peng2002,Mauriz2009}. 
These results suggest a direct link between global mirror symmetry and PTRs.

Recent results, however, report on the presence of PTRs in devices without global mirror symmetry \cite{Lu2005,Nava2009,Grigoriev2010,Zhukovsky2010}, indicating that it is a sufficient but not necessary condition for the appearance of perfect transmission. 
In some cases, the occurrence of PTRs in such devices has been attributed to `internal' \cite{Huang2001} or `hidden' \cite{Nava2009} symmetries. 
In Ref.~\cite{Zhukovsky2010}, conditions for the occurrence of PTRs are derived for hybrid periodic-aperiodic photonic devices, and in Ref.~\cite{Hsueh2011_josab-28-11-2584} PTRs arise in a band gap approach, though without explicit reference to the symmetry of the setups.
These works provide significant insight into the resonant scattering processes in (effectively) 1D inhomogeneous media.
However, the link between transmission properties and the spatial symmetries of the underlying scattering structure has not yet been fully understood.

\begin{figure}[t!]
\vspace{-.5cm}
\includegraphics[width=.98\columnwidth]{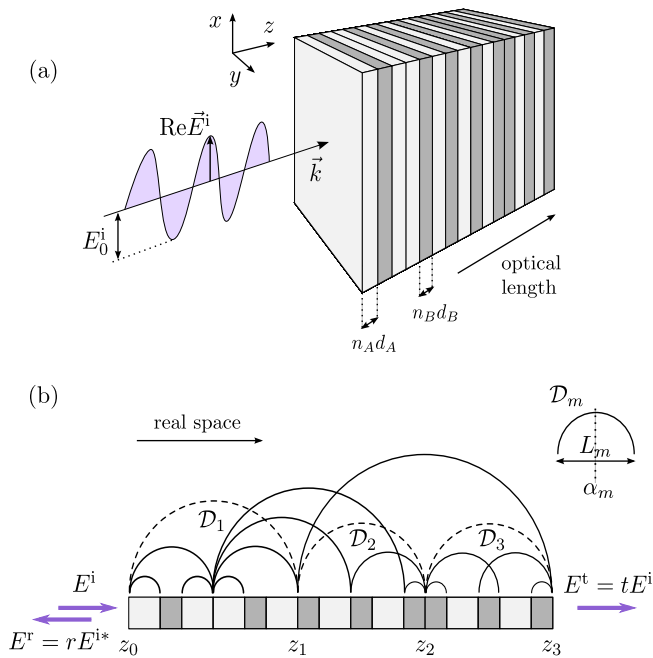}
\vspace{.2cm}
\caption{\label{fig1} (color online) (a) Schematic of an aperiodic multilayer comprised of $16$ planar slabs of materials $A$ (light gray) and $B$ (dark gray), having equal optical thickness $n_{A}d_{A}=n_{B}d_{B}=\lambda_{0}/4$.
The scattered monochromatic plane light wave of stationary electric field amplitude $E$ propagates along the $z$-axis, perpendicularly to the $xy$-plane of the slabs.
(b) 1D cross section of the multilayer in real space, showing its local symmetries. The arcs depict locally symmetric domains ${\mathcal D}_m$ of the device, with lengths $L_m$ and symmetry plane positions $\alpha_m$. For simplicity and figure clarity, only {\it maximal} local symmetries are shown, which are the ones of largest $L_m$ at a given $\alpha_m$ (i.e., any smaller arc, concentric to the ones shown, is also a local symmetry). The selected local symmetry decomposition into $N=3$ domains (dashed arcs) will constitute one of the examples in Sec.~\ref{aperiodic}.}
\end{figure}

A key observation is that, although being globally asymmetric, a system can retain mirror (or, equivalently in one dimension, parity) symmetry within a part of it, thus being {\it locally symmetric} \cite{Kalozoumis2013}.
Indeed, the basic common feature of the above-mentioned asymmetric aperiodic setups is that they can be decomposed into mirror symmetric, non-overlapping smaller parts, which cover the entire device; 
we refer to such systems as \textit{completely locally symmetric}.
Depending on the setup, there can exist a multitude of local symmetry decompositions at different scales and with different symmetry axes.
For example, Fig.~\ref{fig1}(a) depicts a photonic multilayer setup, whose {\it maximal} local symmetries (i.e., of largest range around a given axis) are indicated by arcs in Fig.~\ref{fig1}(b); traversing the setup from left to right along different combinations of arcs, including the non-maximal ones (not shown), yields several different local symmetry decompositions. The question which arises is whether--and in what way--such local symmetry decompositions are related to perfect resonant transmission. 

In the present work we provide a natural classification of PTRs occurring in locally symmetric photonic media, based on the collective contribution of domain-wise invariant, field-dependent quantities characterizing the domains of local symmetry.
This is done by extending the formalism of local parity (LP) symmetry developed in Ref.~\cite{Kalozoumis2013} for 1D quantum scattering, applied here to scattering of classical electromagnetic waves.
The analysis allows for a geometrical representation of three different types of resonances, which is valid for generic 1D variations of the refractive index, and subject only to the restriction to complete local symmetry.
We focus here on piece-wise constant refractive index, which models the intesively studied photonic multilayers, and give an alternative explanation to the origin of PTRs reported for such systems, in terms of local symmetries.

The paper is organized as follows. 
In Sec.~\ref{local_parity} we extend the LP formalism \cite{Kalozoumis2013} to scattering of classical electromagnetic waves in a medium with 1D inhomogeneity, and derive corresponding locally (piecewise) invariant quantities. 
In Sec.~\ref{classification} we use these invariants to classify the possible  PTRs occurring in a completely LP symmetric system within a geometrical representation. In Sec.~\ref{aperiodic}  we apply the LP approach to classify and construct PTRs in aperiodic photonic multilayers. In Sec.~\ref{discussion} we discuss the relation of the LP based classification scheme to other approaches occurring in the literature. Finally, Sec.~\ref{conclusions} summarizes our conclusions.

\section{Local symmetries and invariants \label{local_parity}} 

Before performing the analysis and classification of PTRs in aperiodic photonic multilayers, we will here adapt and further develop the concepts introduced in Ref.~\cite{Kalozoumis2013}, for the case of classical electromagnetic waves.
The electric component of a monochromatic plane light wave of frequency $\omega$ obeys the equation \cite{Joannopoulos2007}
\begin{equation}
\nabla \times \nabla \times \vec{E}(\vec{r},t) =\left( \frac{\omega^2}{c^2} \right) n^2(\vec{r}) \vec{E}(\vec{r},t),
\label{eq:elfi}
\end{equation}
where $n(\vec r)$ is the spatially dependent refractive index.

We consider light propagation in a mixed dielectric medium consisting of regions with different (lossless and dispersionless) dielectric materials which are homogeneous in the $xy$-plane, so that the refractive index varies only in the $z$-direction, $n(\vec{r})=n(z)$.
Further, we restrict the wave to normal incidence on the $xy$-plane, so that it propagates everywhere along the $z$-axis, and the field can thus be written $\vec{E}(\vec{r},t) = E(z)e^{-i\omega t}\hat{z}$, where $E(z)$ the complex field amplitude.

The description then effectively becomes 1D and, dropping the time dependence, Eq.~(\ref{eq:elfi}) acquires the (Helmholtz) form
\begin{equation}
\hat{\varOmega}(z,\omega)E(z) = \frac{\omega^2}{c^2}E(z)
\label{eq:helm}
\end{equation}
where, however, the differential operator
\begin{equation}
\hat{\varOmega}(z,\omega) = -\frac{d^2}{dz^2} + \left[1-n^{2}(z)\right]\frac{\omega^2}{c^2}.
\label{eq:omega}
\end{equation}
depends simultaneously on $n(z)$ and $\omega$.

For a homogeneous medium, $n(z) = const.$, Eq.~(\ref{eq:helm}) becomes an ordinary eigenvalue problem in the squared wavenumber $k^2 = (n\omega/c)^2$.
If $n(z)$ varies, as will be the case in the following, Eq.~(\ref{eq:helm}) is equivalent to a stationary scattering Schr\"odinger equation, by treating $\omega$ as a tunable input parameter.
The problem is then solved for the (complex) transmission amplitude $t$ for an incident wave $E^{\rm i} = e^{ikz}$ to transmit through a given region defined by the scatterer (in our case the photonic multilayer, see Fig.~\ref{fig1})
As the parameter $\omega$ is varied, the transmission spectrum $T(\omega)$ will have fundamental differences from the quantum counterpart \cite{Gaponenko2010}, since the `effective potential' $n(z)$ in Eq.~(\ref{eq:helm}) is multiplied by the `energy' $\omega$.
E.g., classical light will always `feel' the presence of the scatterer, whereas a quantum particle becomes gradually insensitive to it at higher energies ($T(\omega \to \infty) \to 1$).

Since the effective scattering problem, Eq.~(\ref{eq:helm}), is 1D and isomorphic to the quantum counterpart, we can apply the LP formalism introduced in Ref.~\cite{Kalozoumis2013}.
Specifically, we consider a completely locally symmetric setup, that is, a setup which can be decomposed (generally in more than one ways) into $N$ subdomains $\mathcal{D}_{m} = [z_{m-1},z_m]$ ($m=1,...,N$) with 
\begin{equation}
 n(z)=n(2\alpha_m -  z), ~~ z \in \mathcal{D}_m ~~ \forall m, 
\label{eq:refind}
\end{equation}
where $\alpha_m$ is the center of $\mathcal{D}_{m}$ (and, hence, the position of the local symmetry plane of the 3D device; see Fig.~\ref{fig1}).

The key concept is now that of a local mirror reflection through the symmetry plane $z=\alpha_m$ of $\mathcal{D}_{m}$, which is equivalent to a local parity transform in our 1D description.
The action of the two LP operators $\hat{\varPi}^{\mathcal{D}_{m}}_{s_m}$ ($s_m=\pm 1$) on the field is defined as the ordinary parity transform $E(z) \rightarrow E(2\alpha_m -  z)$ within the associated subdomain $\mathcal{D}_{m}$, and, up to a sign, as the identity operator outside $\mathcal{D}_{m}$ \cite{Kalozoumis2013}:
\begin{eqnarray}
\hat{\varPi}_{s_{m}}^{\mathcal{D}_{m}} E(z) ~ = ~~ & {\varTheta} \left(\frac{L_{m}}{2} - |z -\alpha_{m}| \right) & E(2\alpha_m-z) \nonumber \\ 
+ ~s_m & {\varTheta} \left(|z -\alpha_m| - \frac{L_m}{2} \right) & E(z),
\label{eq:lp_def}
\end{eqnarray}
where $L_{m} = z_{m} - z_{m-1}$ is the width of the subdomain $\mathcal{D}_{m}$.
The combined action of LP transforms in all $N$ non-overlapping subdomains defines a {\it total} LP  operator
\begin{equation}
\hat{\varPi} = \prod_{m=1}^{N} \hat{\varPi}_{s_m}^{\mathcal{D}_m}, ~~~ s_{m} \in~\{+1,-1\}.
\label{eq:lp_total}
\end{equation}

The property of complete local symmetry of the scatterer medium gives rise to {\it locally invariant} quantities, i.e., $z$-independent within each subdomain $\mathcal{D}_m$ \cite{translation}, which are constructed as follows.
Multiplying Eq.~(\ref{eq:helm}) by $\hat\varPi E(z)$ and subtracting the $\hat\varPi$-transformed result, we obtain, because of the local symmetry of the refractive index, Eq.~(\ref{eq:refind}),
\begin{equation}
 E''(z)\hat\varPi E(z) - E(z)\hat\varPi E''(z) = 0,
\label{eq:helmdiff}
\end{equation}
which holds for $s_m=+1$ in Eq.~(\ref{eq:lp_total}) for the considered decomposition.

Taken separately in each domain $\mathcal{D}_m$, Eq.~(\ref{eq:helmdiff}) has the form of a total derivative, and can be integrated to give the complex locally invariant quantities
\begin{equation}
E(2 \alpha_{m} - z) E'(z) + E(z) E'(2 \alpha_{m} - z) \equiv Q_{m},
\label{eq:conserved_q}
\end{equation}
where $z \in \mathcal{D}_m$, with $m=1,2,...,N$.
Each $Q_m(\omega)$ can be regarded as a two-point (non-local) `current' which is constant within the corresponding region $\mathcal{D}_m = [z_{m-1}, z_m]$, for {\it any} profile of the field within $\mathcal{D}_m$.
Thereby, the $N$ (generally different) quantities $Q_m$ encode the local symmetry of the scatterer on the level of the field.
The values of the $Q_m$ depend on the considered local symmetry decomposition and on the input frequency $\omega$.
As we will see, in the case of a resonant frequency, they enable the classification of the corresponding field configurations in terms of local symmetries.

To this aim, we evaluate Eqs.~(\ref{eq:conserved_q}) at the planes $z=z_m$, and write them in the form
\begin{equation}
\frac{E'(z_{m-1})}{E(z_{m-1})} + \frac{E'(z_{m})}{E(z_{m})} = \mathcal{V}_m, ~~m=1,2,...,N
\label{eq:conserved_q_divided} 
\end{equation}
where the scaled currents
\begin{equation}
 \mathcal{V}_m \equiv \frac{Q_m}{E(z_{m-1})~E{(z_{m})}}
\end{equation}
characterize the subdomains $\mathcal{D}_m$ at a given $\omega$, involving information only from the field at their boundaries.
We now sum the $N$ Eqs.~(\ref{eq:conserved_q_divided}) with alternating signs $(-1)^m$, yielding
\begin{equation}
\frac{E'(z_0)}{E(z_0)} - (-1)^{N}~ \frac{E'(z_N)}{E(z_N)} = \sum_{m=1}^{N} (-1)^{m-1} \mathcal{V}_m ~ \equiv ~\mathcal{L}
\label{eq:conserved_q_divided_sum} 
\end{equation}
which depends only on the field at the scatterer's {\it global} boundaries $z_0,z_N$, whose norm will in turn determine the occurrence of PTRs.
Therefore, it is convenient to write the electric field $E$ in polar representation, $E(z)= |E(z)|e^{i\varphi(z)} \equiv E_0(z)e^{i\varphi(z)}$, so that Eq.~(\ref{eq:conserved_q_divided_sum}) becomes
\begin{align}
\mathcal{L} = i \left[\varphi'(z_0) - (-1)^{N} \varphi'(z_N)\right] ~~~~~~~~~~~~~~~ \nonumber \\ + \left[ \frac{E_0'(z_0)}{E_0(z_0)} - (-1)^{N} \frac{E_0'(z_N)}{E_0(z_N)} \right]
\label{eq:conserved_q_module} 
\end{align} 
The global quantity $\mathcal{L}$, together with the values of the individual local quantities $\mathcal{V}_m$ in Eq.~(\ref{eq:conserved_q_divided_sum}), can be utilized to classify the scattering states of the system, as we will show next.

\section{Perfect transmission in locally symmetric optical media}

The transmission coefficient $T$ in a photonic scattering setup is defined as the ratio of transmitted to incident light intensity, in our present setting $T = (E^{\rm t}_0/E^{\rm i}_0)^2 = |t|^2$ (see Fig.~\ref{fig1}), where $t$ is the (complex) transmission amplitude.
In the lossless medium we consider, the reflection coefficient is $R = |r|^2 = 1 - T$ due to energy conservation.

As the incident wave propagates through a composite scatterer medium with a varying refractive index, it is multiply scattered and the counterpopagating waves interfere into the stationary scattering state.
Although the various parts of the medium may all exhibit finite reflection, the interference at {\it resonant} frequencies $\omega$ is such that peaks appear in the transmission spectrum $T(\omega)$.

Isolated resonances in a globally symmetric device are typically perfectly transmitting ($T=1$), while  an asymmetric device is usually associated with finite total reflection ($T<1$).
There are, however, cases in which PTRs occur even if the scattering medium is globally asymmetric, and these we will now classify in terms of the quantities in Sec.~\ref{local_parity}.
As previously, we consider devices which are completely locally symmetric, i.e., exactly decomposable into $N$ symmetric units, which is indeed the case for the vast majority of setups used in the (theoretical or experimental) literature.

\subsection{Classification of PTRs \label{classification}}

For perfect transmission, $T=1$, the field magnitude at the global boundaries of the device is $E_0(z\leqslant z_0) = E_0(z \geqslant z_N) = 1$ (having chosen, without loss of generality, a unit amplitude incident wave); 
and due to continuity of $E'(z)$, we also have $E'_0(z_0) = E'_0(z_N) = 0$.
Therefore, at a PTR the second term in Eq.~(\ref{eq:conserved_q_module}) vanishes, and we obtain
\begin{equation}
 \mathcal{L} = iJ[1 - (-1)^N] = \begin{cases} ~0, & N~{\rm even} \\
~2ik,  & N~{\rm odd} \end{cases}
\label{eq:even_odd_sum}
\end{equation}
where $k = \varphi'(z_0) = \varphi'(z_N) = n_0\omega/c$ is the wave number in the ambient medium of refractive index $n_0$.
The quantity $J \equiv E_0^2(z) \varphi'(z) = kT$ is the scaled energy density current (or 1D Poynting vector) $S=\frac{c}{8\pi k}J$ \cite{Chen1987}--in analogy to the probability density current for a quantum matter wave--which is {\it globally} invariant, as opposed to the locally invariant $Q$.

\begin{figure}[t!]
\vspace{.1cm}
\includegraphics[width=.9\columnwidth]{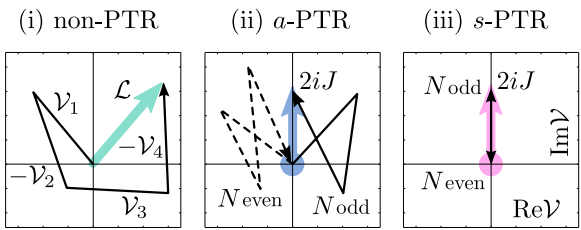}
\vspace{.2cm}
\caption{\label{fig2} (Color online) Geometric representation of scattering in locally symmetric media. The scaled invariants $\mathcal{V}_m$ characterizing each symmetric subdomain $\mathcal{D}_m$ are represented by vectors (thin black lines) and added `head-to-tail' for increasing $m$, with sign $(-1)^{m-1}$ (see text), yielding $\mathcal{L}$ (thick colored lines).
(i) For any non-PTR, the trajectory is open with arbitrary end $\neq 2ik$.
(ii,iii) For any PTR, the trajectory formed by the vectors ends at $0$ ($2ik$) for an even $N$ (odd $N$) LP decomposition of the scattering device. For (ii) asymmetric PTRs the trajectory explores the complex plane, while for (iii) symmetric PTRs it oscillates between $0$ and $2ik$. 
}
\end{figure}

The scaled currents $\mathcal{V}_m$, which are summed along the symmetry domains into $\mathcal{L}$ in Eq.~(\ref{eq:conserved_q_divided_sum}), and which we will represent through vectors in the complex plane, can thus be used to distinguish three main symmetry-based cases of scattering, schematically shown in Fig.~\ref{fig2}:

(i) {\it non-PTR}. 
In this case we have $T < 1$, Eq.~(\ref{eq:even_odd_sum}) is not fulfilled and the `vectors' $\mathcal{V}_m$ add up (with alternating signs) to a complex `vector' $\mathcal{L} \neq 0, 2ik$ (see Fig.~\ref{fig2}(i)). 
The sequence of the added `vectors' thus forms an open trajectory in the complex plane.
Note that the `vectors' are added in the order they appear along the $N$ local symmetry subdomains $\mathcal{D}_m$, in `head-to-tail' manner.

(ii) {\it asymmetric PTR}. 
We call asymmetric PTR ($a$-PTR) a $T=1$ stationary light wave whose electric field magnitude $E_{0}(z)$ is {\it not} completely LP symmetric \cite{Kalozoumis2013} along the $z$-axis in our effective 1D setup.
In this case, the $\mathcal{V}_m$ in Eq.~(\ref{eq:conserved_q_divided}) take on ($z$-invariant but) arbitrary values in the $N$ local symmetry domains (with $N \geqslant 2$ \cite{N1}), determined by the considered decomposition.
For even $N$, the sequence of the added `vectors' forms a closed trajectory in the complex plane, starting and ending at the origin (see dashed line in  Fig.~\ref{fig2}(ii)), as seen from Eq,~(\ref{eq:even_odd_sum}).
For odd $N$, the trajectory is open and ends at $2ik$ (or again closes at zero by adding the fixed `vector' $-2ik$).

(iii) {\it symmetric PTR}.
We call symmetric PTR ($s$-PTR) a resonance which resonates with $T_m=1$ in {\it each} subdomain $\mathcal{D}_m$ (where $T_m$ is the transmission coefficient through $\mathcal{D}_m$ alone) of a considered local symmetry decomposition, with {\it completely locally symmetric} field magnitude $E_0(z)$ following these local symmetries (as was shown in Ref.~\cite{Kalozoumis2013}).
Now $E_0=1$ and $\varphi'=k$ at both boundaries of any subdomain $\mathcal{D}_m = [z_{m-1},z_m]$;
therefore, all local invariants {\it align} to the single, `$N$-fold degenerate' value $Q_m=V_m=2ik$.
The trajectory representing $\mathcal{L}$ is thus restricted to the imaginary axis, oscillating between $0$ and $2ik$ in the complex plane, and ending at $0$ ($2ik$) for even (odd) $N$ (see Fig.~\ref{fig2}(iii)).
We thus have the situation that, at any $s$-PTR, the locally invariant two-point current $Q$ is purely imaginary, with norm twice the globally invariant current $J=k$.

In other words, a PTR is classified as an $s$-PTR simply if $E_0(z)$ is completely locally symmetric (then there {\it exists} at least one LP decomposition with $Q_m=2ik$ and symmetric $E_0(z)$ in each subdomain $\mathcal{D}_m$); otherwise, if $E_0(z)$ is not completely locally symmetric (and the $\mathcal{V}_m$ are not restricted to the imaginary axis for {\it any} LP decomposition), then it is an $a$-PTR.

In general, a given (completely locally symmetric) scattering device can exhibit an $a$-PTR hosting a {\it partial} $s$-PTR over some subdomain(s) of the device, at the same resonant frequency.
Then the $Q_m$ will align along the part(s) of the device where the partial $s$-PTR(s) reside(s), and exhibit a mismatch in the remaining part(s), with combinations of the corresponding type of $\mathcal{V}_m$-trajectories in the complex plane. 

Note here that, if $T=1$, there is no LP decomposition for which Eq.~(\ref{eq:even_odd_sum}) is violated, so that a PTR cannot {\it appear} to be a non-PTR (due to `inappropriate' choice of LP domains) within the proposed classification.
Conversely, a non-PTR cannot appear as a PTR, for any LP decomposition, since Eq.~(\ref{eq:even_odd_sum}) holds only for $r=0$, as can be seen by explicit substitution of $E(z_{0})=e^{ikz_{0}}+r~e^{-ikz_{0}}$ and $E(z_{N})=t~e^{-ikz_{N}}$ into Eq.~(\ref{eq:conserved_q_divided_sum}).

\subsection{PTRs in aperiodic photonic multilayers \label{aperiodic}}

Let us now proceed to investigate the manifestation of the above types of PTRs, and their local symmetry classification, in aperiodic photonic multilayer devices, which are widely used in light transmission experiments \cite{Negro2005,Poddubny2010_Physica.E.42.1871,Negro2011,Vardeny2013}.
Such systems are usually modeled by a piecewise constant refractive index, corresponding to a setup of attached two-dimensional slabs of (usually two, but in general also more) different materials (see Fig.~\ref{fig1}(a)).
Aperiodic multilayers are an ideal implementation for the study of the local symmetry concepts introduced, since the model system is analytically tractable and because they exhibit inherent complete local symmetry.
In fact, most multilayer setups can be decomposed in many ways, with different $N$'s and at multiple scales (see Fig.~\ref{fig1}(b)).
In larger systems, binary aperiodic order (in case of multilayers, of two kinds of slabs $A$ and $B$) can be shown to feature local symmetries with arbitrarily large ranges and high density, and with remarkable symmetry axis distributions \cite{Morfonios2013}.

As a first example, which will also demonstrate the classification of resonances proposed in Sec.~\ref{classification}, we consider the setup schematically shown in Fig.~\ref{fig1}, which is composed of $16$ slabs of materials $A$ and $B$ with refractive indices $n_A$ and $n_B$.
The widths of the slabs are $d_A$ and $d_B$, respectively, such that they have equal optical thickness $n_Ad_A=n_Bd_B=\lambda_0/4$ (the so called `quarter-wave condition' \cite{Joannopoulos2007}), where $\lambda_0$ is a central wavelength.
The slabs are concatenated into a composite scatterer represented by the symbolic sequence $ABAABABABABBABAB$.
This setup has been studied in Ref.~\cite{Nava2009} as the concatenation of the fifth generation $ABAABABA$ of the Fibonacci sequence \cite{Nava2009} to its `conjugate' $BABBABAB$ (where the $A$ and $B$ are interchanged).
We choose this particular setup here because, although being globally asymmetric, it exhibits PTRs at multiple frequencies, as shown in Ref.~\cite{Nava2009}, which we will here identify as the cases described in Sec.~\ref{classification}.
Note also that the aforementioned quarter-wave condition is only used in order to reproduce the corresponding transmission spectrum in Ref.~\cite{Nava2009}, and is not a necessary condition for our approach.

\begin{figure}[t!]
\includegraphics[width=\columnwidth]{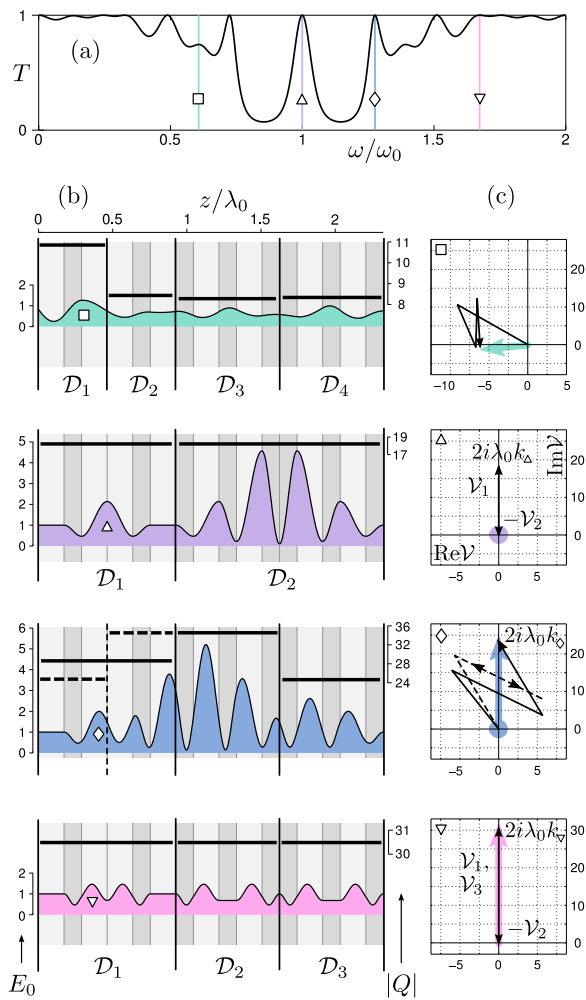}
\caption{\label{fig3} (Color online) 
(a) Transmission coefficient $T$ as function of the scaled frequency $\omega/\omega_0$ for the photonic multilayer shown in Fig.~\ref{fig1} comprised of slabs $A$ and $B$ with refraction indices $n_{A}=2.12$ and  $n_{B}=1.45$. 
(b) Field magnitude $E_0(z)$ across the multilayer at the frequencies marked in (a), corresponding to a non-PTR ($\omega_{\scriptscriptstyle\square} = 0.607 ~\omega_0$), two $s$-PTRs ($\omega_{\vartriangle}=\omega_0$ and $\omega_{\triangledown} = 1.276~\omega_0$), and an $a$-PTR ($\omega_{\scriptstyle\lozenge}$), with scale on the left.
The background shows the slabs $A$ (light gray) and $B$ (dark gray) along the device, and the vertical lines depict the considered decomposition into LP symmetric subdomains of the device.
The $|Q_m|$ for each subdomain $D_m$ are plotted as thick solid lines (scale on the right).
(c) The alternating sum $\mathcal{L}$ (thick colored arrows) of the $\mathcal{V}_m$ for each considered decomposition in (b), represented as a trajectory (thin black arrows) in the complex plane, like in Fig.~\ref{fig2}.
For the $a$-PTR ($\lozenge$), an odd ($N=3$) and an even ($N=4$) decomposition is considered (solid and dashed lines in (b) and (c),respectively).}
\end{figure}

The transmission spectrum $T(\omega)$ of the device is shown in Fig.~\ref{fig3}(a).
It is symmetric around a central frequency $\omega_0 = 2\pi c/ n_0\lambda_0$, where $n_0=n_A$ is the ambient refractive index, as a consequence of the imposed quarter-wave condition \cite{Gellermann1994, Wiersma2005} with $\lambda_0=600~{\rm nm}$.
As we see, there occur several resonances within the plotted range, which indeed are perfectly transmitting.
To demonstrate the manifestation of the presence or absence of local symmetries that distinguish the character of the resonances, the field profiles within the device are plotted in Fig.~\ref{fig3}(b) for selected frequencies (marked in Fig.~\ref{fig3}(a) by the symbols ${\scriptstyle\square}$, ${\vartriangle}$, ${\lozenge}$, ${\triangledown}$), along with the invariants $|Q_m|$ (shown as thick black lines) of the considered decompositions into subdomains $\mathcal{D}_m$.
We consider, for clarity, locally symmetric subdomains containing integer number of slabs, and otherwise of any size (i.e., not restricted to the maximal ones depicted in Fig.~\ref{fig1}(b)). As is clear from the classification scheme in Sec.~\ref{classification}, the decomposition is arbitrary for the manifestation of non- and $a$-PTRs in the complex $\mathcal{V}$-plane, whereas it must be identified as the one matching the $E_0(z)$-profile for an $s$-PTR.
Fig.~\ref{fig3}(c) illustrates the alternating sum $\mathcal{L}$ of each case, represented by the `trajectories' of the $\mathcal{V}_m$ in the complex plane.

The first peak (${\scriptstyle\square}$) is clearly a non-PTR, with $T<1$ and no local symmetries appearing in $E_0(z)$ in the LP subdomains of the device.
This becomes partly evident by the different values of the $|Q_m|$, and fully confirmed by the sum $\mathcal{L}\neq 0,2ik_{\scriptscriptstyle\square}$, which takes on an arbitrary complex value (see thick colored vector in Fig.~\ref{fig3}(c)).
The field of the third resonance (${\lozenge}$) has also no local symmetry, and its $|Q_m|$ vary for the different $\mathcal{D}_m$ in any LP decomposition.
However, as seen in Fig.~\ref{fig3}(c), the different `vectors' $\mathcal{V}_m$ do lead to the value $\mathcal{L}=0$ ($2ik_{\scriptstyle\lozenge}$) for even (odd) LP decomposition, and indeed the wave is perfectly transmitted; a manifestation of an $a$-PTR.
Note that, for the even decomposition ($N=4$, dashed lines in Fig.~\ref{fig3}(b) and (c)), respective `outer' and `inner' local invariants coincide, $\mathcal{V}_1 = \mathcal{V}_4$ and $\mathcal{V}_2 = \mathcal{V}_3$, so that the $\mathcal{L}$-trajectory consists of three points only. 

Finally, the second and fourth selected resonances (${\vartriangle}$, ${\triangledown}$) demonstrate the occurrence of $s$-PTRs, with aligned $|Q_m|=2J=2k_{\vartriangle,\triangledown}$ along the multilayer.
We here see that, depending on the local symmetry decomposition of $s$-PTRs, enhanced localization characteristics can arise within the aperiodic medium: E.g., the central resonance at $\omega_{\vartriangle}$ is strongly localized in $\mathcal{D}_2$, in contrast to the one at $\omega_{\triangledown}$ which is rather delocalized, as seen from the field profiles (Fig.~\ref{fig3}(b)) or anticipated from the resonant widths (Fig.~\ref{fig3}(a); the sharper the resonance, the stronger the localization).

Apart from the $s$-PTRs at $\omega_{\vartriangle}$, at $\omega_{\triangledown}$ and its mirror symmetric frequency $\omega = 2 \omega_{\vartriangle}$ - $\omega_{\triangledown}$, all other PTRs in the plotted spectrum (which has period $2\omega_0$) are $a$-PTRs, with corresponding characteristics in the complex $\mathcal{V}$-plane.
Note that, for the chosen parameters, the left half $ABAABABA$ of the setup does not feature PTRs, as shown in Ref.~\cite{Nava2009}.
It does, however, possess local symmetries, and a suitable tuning of the $d_{A,B}$ and $n_{A,B}$ would render it transparent at certain frequencies (in the form of $a$- or $s$-PTRs).
However, there are then less available LP decompositions, so that the occurrence of {\it multiple} PTRs is relatively limited compared to the present multilayer.
The simple but still representative example studied here clearly illustrates how insight into the properties of resonant waves in aperiodic multilayers is gained by their local symmetry analysis.

\begin{figure}[t!]
\centering
\includegraphics[width=\columnwidth]{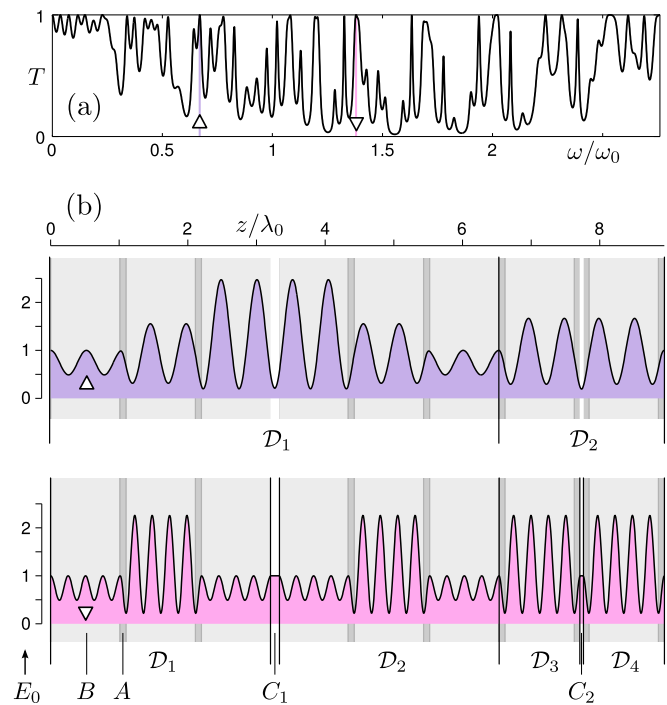}
\vspace{.2cm}
\caption{\label{fig4} (Color online) 
(a) Transmission spectrum around a central frequency $\omega_0=2\pi c/\lambda_0$, with $\lambda_0 = 700~{\rm nm}$, of a photonic multilayer consisting of two different kinds of slabs $A$ ($n_{A}=2.15$, $d_A=0.08426~\lambda_0$) and $B$ ($n_{B}=1.43$, $d_B=1.01348~\lambda_0$), with two intervening gaps $C_1$ and $C_2$ of the ambient medium ($n_{C_1} = n_{C_2}=1$), of widths $d_{C_1}=0.1146~\lambda_0$ and $d_{C_2}=0.04236~\lambda_0$. 
(b) Field amplitudes at the $s$-PTR frequencies $\omega_{\vartriangle}=0.67~\omega_0$ and $\omega_{\triangledown}=1.38~\omega_0$ marked in (a).
The background shows the media $A$ (light gray), $B$ (dark gray) and $C$ (white) along the device, and vertical lines distinguish the considered LP decompositions into resonators at each $s$-PTR.
Note that for $\omega_{\vartriangle}$ the gaps $C$ are parts of the resonators, while for $\omega_{\triangledown}$ they are not (see text).
}
\end{figure}

In a second example, we will show how parametric tuning can indeed enable the construction of $s$-PTRs in a photonic multilayer at {\it prescribed} energies, making use of its local symmetries.
We choose the setup $BABABC_1BABABABAC_2ABA$ shown in Fig.~\ref{fig4}, which is a slight geometrical modification of a multilayer studied in Ref.~\cite{Zhukovsky2010}, where two gaps $C_1$ and $C_2$ of the ambient medium (here vacuum, $n_{C_1}=n_{C_2}=1$) of different widths $d_{C_1}$ and $d_{C_2}$ have been inserted.
Without the gaps, this structure features a single PTR over a wide frequency range \cite{Zhukovsky2010}, and we will now demonstrate the occurrence of two PTRs in the modified setup.
To produce the PTRs we follow the `construction principle' introduced in Ref.~\cite{Kalozoumis2013} (for a formally equivalent quantum mechanical setting), which is here described in the Appendix along with its application to the considered setup.
Essentially, the local symmetries of the device are exploited to reduce the space of available parameters (in the present case the $d_{A,B,C_{1,2}}$ and $n_{A,B,C_{1,2}}$), to be determined from a set of coupled transfer matrix equations for different LP decompositions.
On the other hand, the parameter space must initially be sufficiently large in order to achieve the formation of PTRs at desired frequencies; this is ensured here by the inclusion of the gaps $C_{1,2}$.
Note that, without referring to the decomposition into locally symmetric subdomains of the setup, called `resonators' in Ref.~\cite{Kalozoumis2013}, there is no obvious way to control the frequencies where $s$-PTRs would occur. 

As seen in Fig.~\ref{fig4}(b), we consider two decompositions of the setup:\\ $~~~~~~~~~~~~~~$(${\vartriangle}$) $BABABC_1BABAB|ABAC_2ABA$, \\
consisting of two resonators, and \\ 
$~~~~~~~~~~~~$(${\triangledown}$) $BABAB|C_1|BABAB|ABA|C_2|ABA$, \\ where the two gaps $C_{1,2}$ intervene between four resonators (the symbol $|$ simply indicates the decomposition).
Fig.~\ref{fig4}(a) shows the transmission spectrum of the setup, where the PTRs at two prescribed frequencies are marked correspondingly (${\vartriangle}$ and ${\triangledown}$).
The spectrum is no longer symmetric around $\omega_0$ as in Fig.~\ref{fig3}(a), since we have relaxed the quarter-wave condition used previously.
Although there are many other resonant frequencies, the two marked are truly the only ones with exactly $T=1$, within the plotted range.
Further, they are of $s$-PTR type, as can be anticipated from the field profiles in Fig.~\ref{fig4}(b):
indeed, they follow the local symmetries of the device, according to the indicated resonator decompositions.
For the first PTR (${\vartriangle}$), the gaps $C_{1,2}$ are part of the two resonators, so that the widths $d_{C_{1,2}}$ are determined by the above-mentioned construction principle.
In contrast, for the second PTR (${\triangledown}$), the transparency of the device is independent of $d_{C_{{1,2}}}$, since the gaps are present only {\it between} the considered resonators.
The wave thus propagates only in forward direction within the gaps, as is evident from the corresponding plateaus of the field ($E_0(z)=1$ along the gaps in Fig.~\ref{fig4}(b)(${\triangledown}$)).

Note that, if we set $d_{C_{1}}= d_{C_{2}} = 0$, the second PTR at $\omega_{\triangledown}$ would be the equivalent to the single PTR of the unmodified setup in Ref.~\cite{Zhukovsky2010}.
We see here that, by inserting a third type of slab in the multilayer (here the ambient medium itself), the local symmetries of the device can be exploited to design a new PTR, which would not be possible without the modification.
Further, the new resonant field at $\omega_{\vartriangle}$ is localized on a {\it different spatial scale}, as seen in Fig.~\ref{fig4}(b); E.g., the fields within the respective $\mathcal{D}_1$ at $\omega_{\vartriangle}$ and $\omega_{\triangledown}$ have similar profile, but the latter is `squeezed' to half the range.
Since the PTR at $\omega_{\triangledown}$
is invariant with respect to the gap widths $d_{C_{1,2}}$, we thus see that the LP construction principle can be utilized for the flexible spatial design of resonantly transparent multilayer devices.

\subsection{Relation to alternative approaches \label{discussion}}

Having demonstrated how the concept of local symmetries enable the `geometric' classification of resonant scattering for photonic multilayers and the construction of PTRs, we now briefly discuss it in relation to other approaches to PTRs.
As already mentioned in Sec.~\ref{introduction}, PTRs have received considerable attention, in particular for photonic multilayers which can be realized with high accuracy and very efficient transmission characteristics.
There is, indeed, a number of different theoretical approaches dealing with the occurrence of PTRs in model systems.
In Ref.~\cite{Hsueh2011_josab-28-11-2584}, e.g., PTRs arise from the intersection or touching of transmission bands in a periodic extension of a given (aperiodic) device with the variation of some parameter (like the slab widths).
Another approach is to identify PTRs via the phase accumulated by the counterpropagating waves within the scatterer, which gives an interpretation of the vanishing reflection \cite{Chuprikov1993}.
A particularly relevant approach is given by Zhukovsky in Ref.~\cite{Zhukovsky2010}, where the PTRs of combined photonic multilayers are classified with respect to the transmissions of their parts.
Most of these works, however, focus more on the {\it conditions} for the occurrence of PTRs and less on the understanding of their {\it origin} from fundamental principles.
The latter is captured here, on the level of the field magnitude, within the classification of PTRs on (local) symmetry grounds.

More specifically, in Ref.~\cite{Zhukovsky2010} a given photonic multilayer is composed of a left and a right part covering two domains $\tilde{\mathcal{D}}_1$ and $\tilde{\mathcal{D}}_2$, respectively (just like the domains $\mathcal{D}_1$ and $\mathcal{D}_2$ in Fig.~\ref{fig3}(b)(${\vartriangle}$)).
The tilde here indicates that these domains, although having their boundaries at the interfaces between slabs, do {\it not} necessarily contain symmetric refractive index $n(z)$.
Using the Airy formulas \cite{Sun1991} for the transmission of composite 1D systems, it can be shown that the total setup is perfectly transmitting, $T=1$, if one the following conditions is fulfilled \cite{Zhukovsky2010}:
\begin{subequations}
\label{zhuk12}
\begin{align}
\label{zhuk1} &T_1=T_2=1 ~; \\
\label{zhuk2} &T_1=T_2 \neq 1~,~~ (\overline\phi_1 + \phi_2){\rm mod}2\pi = 0, 
\end{align}
\end{subequations}
where $T_{1(2)}$ are the transmission coefficients through the subdomain $\tilde{\mathcal{D}}_1$ ($\tilde{\mathcal{D}}_2$) and $\overline{\phi}_1$ ($\phi_2$) are the phases of the reflection amplitude $\overline{r}_1 = \sqrt{R_1}e^{i\overline\phi_1}$ ($r_2 = \sqrt{R_2}e^{i\phi_2}$). 
The bar indicates that $\tilde{\mathcal{D}}_1$ is traversed in the opposite ($-z$) direction.
Note here that, although simple and appealing, the classification of PTRs through Eqs.~(\ref{zhuk12}) makes no reference to the underlying structure of the field within the multilayer.
Moreover, it is restricted to decompositions of the scatterer into $N=2$ domains, or to pairs of neighboring domains in an $N>2$ decomposition.

For $N=2$ domains in the decomposition of a device, it is clear that the $s$-PTRs defined in Sec.~\ref{classification}, which are locally symmetric within the two domains, correspond to the first case above, Eq.~(\ref{zhuk1}).
The $a$-PTRs then necessarily correspond to the second case, Eq.~(\ref{zhuk2}), whose fulfillment depends, however, on the particular decomposition.
For instance, the $a$-PTR in Fig.~\ref{fig3}(b)(${\lozenge}$) belongs to this case, but only if the multilayer is decomposed into two parts $\mathcal{D}_1$ and $\mathcal{D}_2$ as $ABAABA|BABABBABAB$.

For $N>2$, there is an `overlap' of the two cases in the classification of Eqs.~(\ref{zhuk12}), in the sense that a given PTR can belong simultaneously to both.
Assume, for instance, that the left part of the multilayer $\tilde{\mathcal{D}}_1\tilde{\mathcal{D}}_2$ considered above is composed of two smaller parts as $\tilde{\mathcal{D}}_1 = \tilde{\mathcal{D}}_1^a\tilde{\mathcal{D}}_1^b$, for which the second condition, Eq.~(\ref{zhuk2}), holds ($T_1^a \neq T_1^b$).
Then, a PTR of the {\it total} multilayer fulfills Eq.(\ref{zhuk1}) for the decomposition $\tilde{\mathcal{D}}_1|\tilde{\mathcal{D}}_2$, and Eq.~(\ref{zhuk2}) for the decomposition $\tilde{\mathcal{D}}_1^a|\tilde{\mathcal{D}}_1^b\tilde{\mathcal{D}}_2$;
that is, the PTR belongs to both categories for the same setup and frequency.
As a consequence, Eqs.~(\ref{zhuk12}) will provide the conditions for perfect transmission, but fail to reveal the `true nature' of the resonance.
With the `geometric' classification pursued here, the resonance is unambiguously identified as an $s$- or $a$-PTR, since it will follow the local symmetries of the setup or not, respectively, as explained in Sec.~\ref{classification}.

In other words, for $N>2$ decomposition domains $\tilde{\mathcal{D}}_m$ (not necessarily reflection symmetric) there is no one-to-one mapping between the conditions in Eqs.~\ref{zhuk12} and the $s$- and $a$-PTRs.
The two approaches can, however, be used in complementary fashion:
With a simple local symmetry analysis of the device, the locally invariant $|Q|$-values determine if a PTR is symmetric or asymmetric, and then the potential $a$-PTR parts (i.e., parts in the multilayer with unequal $|Q|$) can be checked for fulfillment of Eq.~(\ref{zhuk2}).

\section{Conclusions} \label{conclusions}

We have developed a classification of resonances occuring in wave scattering through any finite, effectively 1D medium which is globally asymmetric but decomposable into locally reflection (or parity) symmetric units.
This was done by exdending the principle of locally invariant currents within the local parity (LP) formalism introduced in Ref.~\cite{Kalozoumis2013}, here applied to classical light scattering.
Emphasizing on the manifestation of perfectly transmitting resonances (PTRs), we used these local invariants, which are determined by the field at the symmetry subdomain interfaces, to achieve a geometrical representation of three classes of scattering states in the complex plane:
(i) non-PTRs, for which the local invariants are different among the subdomains of a considered decomposition of the scattering device, forming an arbitrary trajectory in the complex plane, 
(ii) asymmetric PTRs ($a$-PTRs), where the invariants still differ but with trajectory closing at the origin or ending at $2iJ$, where $J$ is the globally invariant energy density current, and
(iii) symmetric PTRs ($s$-PTRs), where all local invariants align at $2iJ$, with the field magnitude following the local symmetries of the device.

Focusing on optical transmission through aperiodic multilayers of varying refractive index, we demonstrated the local symmetry classification of resonances in representative setups, giving insight into the structure and origin of each type of PTR.
Further, it was shown how simultaneous local symmetries at different scales can be utilized to design aperiodic photonic multilayers with PTRs at prescribed frequencies.
We finally discussed the complementary relation of the present distinction between $a$- and $s$-PTRs to alternative approaches to perfect transmission.
In conclusion, the proposed classification provides an unabmiguous distinction between resonances based on fundamental (local) symmetry principles, and may add to the deeper understanding of the mechanisms underlying resonant scattering in complex systems.

\section{Acknowledgments} \label{acknowledgments} 
This research has been co-financed by the  European Union (European Social Fund - ESF) and Greek national funds through the Operational Program 
``Education and Lifelong Learning" of the National Strategic Reference Framework (NSRF) - Research Funding Program: Heracleitus
II. Investing in knowledge society through the European Social Fund. Further financial support by the Greek Scholarship Foundation IKY in the framework of an exchange program with 
Germany (IKYDA) is also acknowledged. The authors would like to thank Prof. N. Stefanou for illuminating discussions. P.S. acknowledges the hospitality and many fruitful discussions at the Institute for Theoretical Atomic, Molecular and Optical Physics at the Harvard Smithsonian Center for Astrophysics in Cambridge, USA.

\appendix*
\section{Construction of $s$-PTRs} \label{appendix} 

It is here explained in detail, how the concept of LP symmetry is utilized to construct (multiple) $s$-PTRs at preselected frequencies. 
Consider a photonic multilayer setup comprised of $N_{\mathcal{S}}$ homogeneous, plane slabs, like in Fig.~\ref{fig1}.
Let us now assume that the total setup can be decomposed in $N_{\mathcal{D}}$ different ways into locally symmetric domains (considering, for clarity, the slabs themselves as smallest building blocks), so that the $i$-th decomposition has $N^{(i)} \leqslant N_{\mathcal{S}}$ ($i=1,2,...,N_{\mathcal{D}}$) local symmetry subdomains ${\mathcal{D}}^{(i)}_m$ ($m=1,2,...,N^{(i)}$).
In turn, the $m$-th subdomain, which we call `resonator', contains $N^{(i)}_m$ slabs (so that $\sum_m N^{(i)}_m = N_{\mathcal{S}}$ for any $i$).

The unimodular transfer matrix (TM) connecting the plane wave amplitudes on either side of the $m$-th resonator in the $i$-th decomposition is given by the product (ordered in $l$)
\begin{equation}
\label{appendix1} 
M^{(i)}_m = \left( \begin{array}{cc} w^{(i)}_m & z^{(i)}_m \\ z^{(i)*}_m & w^{(i)*}_m \end{array} \right) = \prod_{l=1}^{N^{(i)}_m} M^{(i)}_{m,l}(\omega;n^{(i)}_{m,l},d^{(i)}_{m,l})
\end{equation}
where $M^{(i)}_{m,l}$ is the TM of the $l$-th slab in this resonator with refraction index $n^{(i)}_{m,l}$ and width $d^{(i)}_{m,l}$.
E.g., for the selected $i$-th (dashed) decomposition in Fig.~\ref{fig1}(b), we have $N^{(i)}=3$, $N^{(i)}_{1,2,3} = 6,5,5$, and $n^{(i)}_{m,l} = n_A$ ($n_B$) for $\{m=1;~l=1,(2),3,4,(5),6\}$ and $\{m=2,3;~l=(1),2,(3),4,(5)\}$.

For an $s$-PTR to occur for the $i$-th decomposition at a selected frequency $\omega_i$, that is, with $T^{(i)}_m = 1$ in each of its subdomains ${\mathcal{D}}^{(i)}_m$, the corresponding TM elements $z^{(i)}_m$ must vanish at $\omega = \omega_i$.
Thus, if we want to construct $s$-PTRs at {\it different} frequencies $\omega_{i_1},\omega_{i_2},...$ for equally many LP decompositions $i_1, i_2,...$ of the {\it same} multilayer setup, the corresponding TM elements must solve the following system of $N^{(i)}$ algebraic equations:
\begin{equation}
\label{appendix2}
z^{(i)}_m(\omega_i;n^{(i)}_{m,\{l\}},d^{(i)}_{m,\{l\}}) = 0,~~~ m=1,2,...,,N^{(i)}
\end{equation}
for all selected $i = i_1,i_2,...$ {\it simultaneously}, where $\{l\}$ denotes the set of slabs in the $m$-th resonator.
With the $\omega_i$ fixed at desired values, these $(N^{(i_1)}+N^{(i_2)}+...)$ algebraic equations determine equally many slab parameters (widths and refractive indices), while the remaining ones are set to appropriate (physically relevant) values.

The solution of the system thus provides us with a multilayer setup with PTRs at the prescribed frequencies.
Note that, if the materials and widths of different slabs are chosen equal, as is usually the case, then the number of parameters to be determined is accordingly reduced.
Therefore, to obtain an acceptable combination of slab parameters, a sufficiently large flexibility is needed in the {\it geometry} of the setup, that is, the number of slabs and their order in the multilayer.
The key role of the local symmetries in the above procedure then lies in the reduction of the space of combinations of decompositions for which to establish perfect transmission:
If we had not considered LP symmetric decompositions, then there would be vastly many combinations of decompositions for which to seek a common solution.
Those are now restricted by considering only locally symmetric ones, relying on the one-to-one correspondence between $s$-PTRs and LP symmetry.

The `construction principle' described above is implemented to produce the two $s$-PTRs in the second example of Sec.~\ref{aperiodic}, shown in Fig.~\ref{fig4}, as follows. 
As indicated in Fig.~\ref{fig4}(b), the multilayer setup consists of two kinds of slabs $A$, $B$, and two intervening gaps $C_1$, $C_2$ of the ambient medium (vacuum), with refractive indices and widths $n_A$, $n_B$ and $d_A$, $d_B$, and $n_{C_1} = n_{C_2} \equiv 1$ and $d_{C_1}$, $d_{C_2}$, respectively.
The setup is decomposed into resonators in two ways, labeled $i=\vartriangle, \triangledown$ (see Sec.~\ref{aperiodic}), in order to produce two corresponding PTRs at the frequencies $\omega_{\vartriangle},~\omega_{\triangledown}$,
\begin{center}
${\mathcal{D}}^{(\vartriangle)}_1 {\mathcal{D}}^{(\vartriangle)}_2$ ~ and ~ 
${\mathcal{D}}^{(\triangledown)}_1 C_1 {\mathcal{D}}^{(\triangledown)}_2 {\mathcal{D}}^{(\triangledown)}_3 C_2 {\mathcal{D}}^{(\triangledown)}_4$. 
\end{center}
Note that, in the first case ($\vartriangle$), the gaps $C_{1,2}$ are part of the considered resonators, and so their widths $d_{C_{1,2}}$ are relevant for the resonance condition.
In the second case ($\vartriangle$), they are not part of the resonators, so that an $s$-PTR in this decomposition will be retained irrespectively of the gap widths (since there is no reflection along the gaps at any frequency).

According to the procedure described above, we firstly compute the TM elements (six in total) of each resonator in both decompositions as a function of the setup parameters and the selected frequencies,
\begin{center}
$z^{(\vartriangle)}_{1(2)}(\omega_{\vartriangle};n_{A,B},d_{A,B,C_{1(2)}})$, ~
$z^{(\triangledown)}_{1,2,3,4}(\omega_{\triangledown};n_{A,B},d_{A,B})$.
\end{center}
The conditions for $s$-PTRs at the desired frequencies $\omega_{\vartriangle}=0.67$ and $\omega_{\triangledown}=1.38$, Eqs.~(\ref{appendix2}), applied on each of the TM elements above, then yields a system of six algebraic equations which determines the remaining six parameters $n_{A,B},d_{A,B,C_{1,2}}$ of the setup.
We here solved for the gap widths $d_{C_{1,2}}$ having preselected $\omega_{\vartriangle,\triangledown}$; 
conversely, we could vary the $d_{C_{1,2}}$ as input parameters and solve the system for the (unknown) resonant frequencies.

Note here that, under the restriction of having two kinds of slabs $A$ and $B$, the two intervening gaps $C_{1,2}$ are chosen as a minimal geometric deviation from the corresponding setup used in Ref.~\cite{Zhukovsky2010}, in order to get two PTRs at {\it selected} frequencies:
Had the gaps been absent, then all six parameters $n_{A,B},d_{A,B},\omega_{\vartriangle,\triangledown}$ would have been determined from the solution (if existent) of six equations, that is, we could neither have selected resonant frequencies nor tuned some parameter(s) to obtain physically acceptable values for the rest.
By introducing more kinds of slabs (i.e., $n$'s and/or $d$'s) in the multilayer, the parameter space can be broadened to obtain further PTR choices.


\begin{thebibliography}{99}

\bibitem{PhysRevLett.91.108101} S. Roche, Phys. Rev. Lett {\bf 91}, 108101 (2003)
\bibitem{Macia2006} E. Maci\'a, Rep. Prog. Phys. {\bf 69}, 397 (2006)
\bibitem{PhysRevB.76.235113} A. Nomata and S. Horie, Phys. Rev. B {\bf 76}, 235113 (2007)

\bibitem{Hsueh2011_josab-28-11-2584} W. J. Hsueh, S. J. Wun, Z. J. Lin, and Y. H. Cheng, J. Opt. Soc. Am. B {\bf 28}, 2584 (2011)
\bibitem{Werchner2009} M. Werchner, M. Schafer, M. Kira, S. W. Koch, J. Sweet, J.D. Olitzky, J. Hendrickson, B. C. Richards, G. Khitrova, H. M. Gibbs, A. N. Poddubny, E. L. Ivchenko, M. Voronov and M. Wegener, Opt. Express {\bf 17}, 6813 (2009). 
\bibitem{Macia2012} E. Maci\'a, Rep. Prog. Phys. {\bf 75}, 036502 (2012).


\bibitem{Hladky2013} A. C. Hladky-Hennion, J. O. Vasseur, S. Degraeve, C. Granger, and M. de Billy, J. Appl. Phys {\bf 113}, 154901 (2013).
\bibitem{Hsueh2013} W. J. Hsueh, C. H. Chen, R. Z. Qiu, Phys. Lett. A {\bf 377}, 1378 (2013). 

\bibitem{Kohmoto1987} M. Kohmoto, B. Sutherland, and K. Iguchi, Phys. Rev. Lett. {\bf 58},
2436 (1987).
\bibitem{Gellermann1994} W. Gellermann, M. Kohmoto, B. Sutherland, and P. C. Taylor,
Phys. Rev. Lett. {\bf 72}, 633 (1994).
\bibitem{Macia1998} E. Maci\'a, Appl. Phys. Lett. {\bf 73}, 3330 (1998).

\bibitem{Sun1991} X. Sun and D. L. Jaggard, J. Appl. Phys. {\bf 70}, 2500 (1991). 
\bibitem{Lavrinenko2002} A. V. Lavrinenko, S. V. Zhukovsky, K. S. Sandomirski, and
S. V. Gaponenko, Phys. Rev. E {\bf  65}, 036621 (2002).
\bibitem{Zhukovsky2004} S. V. Zhukovsky, A. V. Lavrinenko, and S. V. Gaponenko,
Europhys. Lett. {\bf 66}, 455 (2004).
\bibitem{Zhukovsky2005} S. V. Zhukovsky and A. V. Lavrinenko, Photon. Nanostruct.
Fundam. Appl. {\bf 3}, 129 (2005).
\bibitem{Chuprikov2006} N. L. Chuprikov, O. D. Spiridonova, J. Phys. A {\bf39}, L559 (2006).

\bibitem{Zhukovsky2008} S. V. Zhukovsky and S. V. Gaponenko, Phys. Rev. E {\bf 77}, 046602 (2008).
\bibitem{Nava2009} R. Nava, J. Taguena-Martinez, J. A. del Rio, and G. G. Naumis,
J. Phys.: Condens. Matter {\bf 21}, 155901 (2009).
\bibitem{Zhukovsky2010} S. V. Zhukovsky, Phys. Rev. {\bf A 81}, 053808 (2010).

\bibitem{Huang1999} X. Huang, Y.Wang, and C. Gong, J. Phys.: Condens. Matter {\bf 11},
7645 (1999).
\bibitem{Huang2001} X. Q. Huang, S. S. Jiang, R. W. Peng, and A. Hu, Phys. Rev. B {\bf  63}, 245104 (2001).
\bibitem{Peng2002} R. W. Peng, X. Q. Huang, F. Qiu, Mu Wang, A. Hu, S. S. Jiang, and M. Mazzer, Appl. Phys. Lett. {\bf 80}, 3063 (2002).
\bibitem{Mauriz2009} P. W. Mauriz, M. S. Vasconcelos, and E. L. Albuquerque, Phys. Lett. A {\bf 373}, 496 (2009).

\bibitem{Lu2005} Ye Lu, R. W. Peng, Z. Wang, Z. H. Tang, X. Q. Huang, Mu Wang, Y. Qiu, A. Hu, S. S. Jiang, and D. Feng, J. Appl. Phys. {\bf 97}, 123106 (2005).
\bibitem{Grigoriev2010} V. Grigoriev and F. Biancalana, New J. Phys. {\bf 12} 053041 (2010).

\bibitem{Kalozoumis2013} P. A. Kalozoumis, C. Morfonios, F. K. Diakonos, and P. Schmelcher, Phys. Rev. A {\bf 87}, 032113 (2013).

\bibitem{Joannopoulos2007} J. D. Joannopoulos, R. D. Meade, and J. N. Winn, {\it Photonic Crystals: Molding the Flow of Light}, Princeton University Press, Princeton, NJ (2007).

\bibitem{Gaponenko2010} Sergey V. Gaponenko, {\it Introduction to Nanophotonics}, Cambridge University Press, Cambridge, UK (2010).

\bibitem{translation} `Local translation invariance' of a complex function $f(z)$ within a subdomain $\mathcal{D} = [z_a,z_b] \subset \mathbb{R}$ of continuous space is defined here as follows: For any $z \in \mathcal{D}$ and $\delta \leqslant z_b - z$, $f(z+\delta) = f(z)$. In other words, $f(z) = c_\mathcal{D}$ is $z$-independent within subdomain $\mathcal{D}$, and can have values $\neq c_\mathcal{D}$ outside $\mathcal{D}$.

\bibitem{Chen1987} W. Chen and D. L. Mills, Phys. Rev. B {\bf 35}, 524 (1987)

\bibitem{N1} For $N=1$, any PTR is necessarily LP symmetric, as shown in Ref.~\cite{Kalozoumis2013}.

\bibitem{Negro2005} L. Dal Negro, J. H. Yi, V. Nguyen, Y Yi, J. Michel, and L. C. Kimerling, Appl. Phys. Lett. {\bf 86}, 261 (2005). 
\bibitem{Poddubny2010_Physica.E.42.1871} A. Poddubny and E. Ivchenko, Physica E {\bf 42},  1871  (2010).
\bibitem{Negro2011} L. Dal Negro and S. V. Boriskina, Las. Phot. Rev. {\bf 6}, 178 (2012).  
\bibitem{Vardeny2013} Z. V. Vardeny, A. Nahata, and A. Agrawal, Nat. Phot. {\bf 7}, 177 (2013). 

\bibitem{Morfonios2013} C. Morfonios,  P. Schmelcher, P. A. Kalozoumis and F. K. Diakonos, arXiv:1307.1838.

\bibitem{Wiersma2005} D. S. Wiersma {\it et al.}, J. Opt. A: Pure Appl Opt. 7 , S190 (2005).

\bibitem{Chuprikov1993} G. F. Karavaev and N.L. Chuprikov,  Russ. Phys. J. {\bf 36}, 228 (1993); Russ. Phys. J. {\bf 36}, 749 (1993).

\end{thebibliography}
\end{document}